\documentstyle[12pt,a4wide]{article}
\newcommand{\nl}{ {\hfill \break} }
\newcommand{\np}{ {\newpage } }

\newcommand{\R}{ \mbox{\rm I$\!$R} }

\newcommand{\sign}{ \mbox{\rm sign} }

\newcommand{\artanh}{ \mbox{\rm artanh} }
\newcommand{\arcoth}{ \mbox{\rm arcoth} }

\newcommand{\aBlLP}{1}
\newcommand{\bBlLP}{2}
\newcommand{\Iv }{{3}}
\newcommand{\aZh}{{4}}
\newcommand{\bZh}{{5}}
\newcommand{\HP }{{6}}
\newcommand{\Bl   }{7}
\newcommand{\IMZ}{{8}}
\newcommand{\bRa}{{9}}
\newcommand{\Pa }{{10}}
\newcommand{\Sch}{{11}}
\newcommand{\Gi }{{12}}
\newcommand{\Ma }{{13}}
\newcommand{\Ga   }{{14}}
\newcommand{\Vi}{{15}}
\newcommand{\aRa}{{16}}
\newcommand{\Ze }{{17}}
\newcommand{\RS }{{18}}
\newcommand{\Xa }{{19}}
\newcommand{\bBlZ }{{20}}

\begin{document}

\vspace{-0.5truecm}

\hfill    {\bf FUB - HEP/94 - 3 v.2}

\bigskip

\bigskip

\vspace{-0.2truecm}

\centerline{\large \bf CLASSICAL AND QUANTUM SOLUTIONS}
\centerline{\large \bf OF CONFORMALLY RELATED }
\centerline{\large \bf MULTIDIMENSIONAL COSMOLOGICAL MODELS}

\vspace{-0.03truecm}

\bigskip

\centerline{\bf \large U. Bleyer\dag, M. Rainer\dag, A. Zhuk{\dag \ddag}}

\vspace{0.36truecm}

\centerline{\dag Gravitationsprojekt, Universit\"at Potsdam}
\centerline{An der Sternwarte 16}
\centerline{D-14482 Potsdam, Germany}

\vspace{0.15truecm}

\centerline{\ddag Fachbereich Physik, Freie Universit\"at Berlin}
\centerline{Arnimallee 14}
\centerline{ D-14195 Berlin, Germany \footnote{Permanent address:
Department of Physics,
University of Odessa,
2 Petra Velikogo,
Odessa 270100, Ukraine}}

\vspace{0.35truecm}

\centerline{\bf Abstract}

\vspace{0.07truecm}

{\small
\noindent
We consider multidimensional universes
$M= \R\times M_1\times \cdots \times M_n$ with $D=1+\sum^n_{i=1} d_i$,
where the $M_i$ of dimension $d_i$ have constant curvature,
being compact for $i>1$.
For Lagrangian models $L(R,\phi)$
on $M$ which depend only on Ricci curvature $R$ and a scalar field
$\phi$, there exists an explicit description of conformal equivalence,
with the minimal coupling model and the conformal coupling model
as distinguished representatives of a conformal class.
For the conformally coupled model we study classical
solutions and their relation to solutions in the equivalent minimally
coupled model. The domains of equivalence are separated by certain
critical values of the scalar field $\phi$. Furthermore the coupling
constant $\xi$ of the coupling between $\phi$ and $R$ is critical at
both, the minimal value $\xi=0$ and the conformal value
$\xi_c=\frac{D-2}{4(D-1)}$.
In different noncritical regions of $\xi$ the solutions behave qualitatively
different.
For vanishing potential of the minimally coupled scalar field we find a
multidimensional generalization of Kasner's solution. Its scale factor
singularity vanishes in the conformal coupling model.
Static internal spaces in the minimal model become dynamical
in the conformal one. The nonsingular conformal solution has a particular
interesting region, where internal spaces shrink while the external space
expands. While the Lorentzian solution relates to a creation of the
universe at finite scale, it Euclidean counterpart is an (instanton)
wormhole.
Solving the Wheeler de Witt equation we obtain the quantum counterparts
to the classical solutions. A real Euclidean quantum wormhole is
obtained in a special case.}



\np
\section{\bf Introduction}
\setcounter{equation}{0}
Recently gravitational models in multidimensional universes
$M= \R\times M_1\times \cdots \times M_n$, with $D=1+\sum^n_{i=1} d_i$,
have received
increasing interest.
The geometry might be minimally coupled to
a spacially homogeneous scalar field $\Phi$ with a potential
$U(\Phi)$.
This class of minisuperspace models is rich
enough to study the relation and the imprint of internal compactified
extra dimensions (like in  Kaluza-Klein models$^{\aBlLP,\bBlLP}$) on
the external space-time.

In order to obtain quantum cosmological solutions,
within the framework of canonical quantum gravity,
the Wheeler de Witt (WdW)
equation has to be solved on the
minisuperspace corresponding
to multidimensional geometry
minimally coupled to
a spacially homogeneous scalar field $\Phi$ with a potential $U(\Phi)$.

In Ref. \Iv\ a criterion of integrability for classes
of multidimensional geometry has been
found by analogy with Toda systems. E.g. when there is only one factor
space, say $M_1$, which is non Ricci flat, the system is integrable.
Furthermore in Refs. \aZh\ and \bZh\ interesting quantum solutions have
been found, including quantum wormholes$^\HP$.

In Sec. 2 the theory for classical multidimensional universes
$M= \R\times M_1\times \cdots \times M_n$ with $D=1+\sum^n_{i=1} d_i$,
is sketched, following the setup in Refs. \bBlLP, \Bl\ and \IMZ,
where we are
mainly interested in the case in which the
$d_i$-dimensional spaces $M_i$ are of constant curvature and the internal
spaces,
with $i>1$, are compact.

A special emphasis is put to compare existing natural time gauges$^\bRa$,
given by the choices of
i) the synchronous time $t_s$ of the universe $M$,
ii) the conformal time $\eta_i$ of a universe with the only spacial factor
$M_i$,
iii) the mean conformal time $\eta$, given differentially as some
scale factor weighted average of $\eta_i$ for all $i$ and
iv) the harmonic time $t_h$, which will be used as specially
convenient in calculations on minisuperspace, since in this gauge
the minisuperspace lapse function is $N\equiv 1$.

For a multidimensional universe $M$
the pure geometrical
Einstein-Hilbert theory with
Gibbons-Hawking boundary term, allowing a description on a minisuperspace,
is minimally coupled to
a spacially homogeneous scalar field $\Phi$ with a potential $U(\Phi)$,
and this Lagrangian model  is equivalent to a new one
on an  enlarged minisuperspace.
The motivation for such a introduction is given by the request
for an inflational cosmology.

In Sec. 3 we examine conformal transformations of Lagrangians
on a $D$-dimensional space-time, first generally and then consider as
example of special interest the conformal transformation between a
model with minimally coupled scalar field and an equivalent conformal model
with a conformally coupled scalar field, thus generalizing
previous results from Refs. \Pa\ and \Sch\, obtained for $n=1$ and $D=4$.
We compare the solutions of the minimal
model to their conformal counterpart.

In the Ricci flat case with vanishing potential in the minimal
coupling model
a generalized Kasner solution is obtained.
In the special case of only statical internal spaces in the minimal model,
we get particularly interesting Lorentzian and Euclidean solutions
in the conformal model.
Internal spaces which
are static in the minimal model show interesting
dynamics against  external space in the conformal model.

While in the minimal model time is harmonic, it is no longer harmonic
in the conformal model.
It is a characteristic feature
that natural time gauges are not preserved under conformal transformation
of Lagrangian models.
The synchronous time pictures of the minimal and conformal coupling models
are calculated.

In Sec. 4 we will investigate
the quantum analogue of the classical
solution for the particular model of Sec. 3
with all $M_i$ Ricci flat, and especially the degenerate
case coresponding classically to static internal spaces
in the minimal (coupling) model. We discuss also the
quantum wave function corresponding to classical conformal model.

Examination of the transition to the Euclidean region
provides in the case of real geometry
a quantum wormhole solution according to the boundary conditions of Ref. \HP.

In the Conclusion we resume the perspective of the present results.

\section{\bf Classical Multidimensional Universes}
\setcounter{equation}{0}
We consider a universe described by a (Pseudo-) Riemannian manifold
$$
M=\R\times M_1 \times\ldots\times M_n,
$$
with first fundamental form
\begin{equation}
g\equiv ds^2 = -e^{2\gamma} dt\otimes dt
     + \sum_{i=1}^{n} a_i^2 \, ds_i^2,
\end{equation}
where    $ a_i=e^{\beta^i} $
is the scale factor of the $d_i$-dimensional space $M_i$.
In the following we assume $M_i$ to be an Einstein space,
i.e. its first fundamental form
\begin{equation}
ds_i^2
=g^{(i)}_{{k}{l}}\,dx^{k}_{(i)} \otimes dx^{l}_{(i)}
\end{equation}
satisfies the equations
\begin{equation}
R^{(i)}_{kl}=\lambda_i g^{(i)}_{kl},
\end{equation}
and hence
\begin{equation}
R^{(i)}=\lambda_i d_i.
\end{equation}
Here the Ricci tensor and scalar are defined as usual by
\begin{equation}
R_{\mu\nu}:=R^\lambda_{\mu\lambda\nu}
\quad{\rm and}\quad
R:=R^\mu_\mu.
\end{equation}
Especially we will keep in mind the interesting subcase where
$M_i$ is of constant curvature. In this case
\begin{equation}
ds_i^2
=\frac{1}{(1+\frac{1}{4}K_i r_i^2)^2}\sum_{k=1}^{d_i}
dx^{k}_{(i)} \otimes dx^{k}_{(i)},
\end{equation}
with radial variable
$r_i=\sqrt{\sum_{k=1}^{d_i}\left(x^k_{(i)}\right)^2}$
and constant
sectional curvature, normalized with $K_i=\pm 1$ for positive and negative
$K_i$ respectively. In the flat case $K_i=0$.
Then the Riemann tensor of $M_i$ is
\begin{equation}
R^{(i)}_{klmn}=K_i(g^{(i)}_{km}g^{(i)}_{ln}-g^{(i)}_{kn}g^{(i)}_{lm}).
\end{equation}
Ricci tensor and scalar are then given by Eq. (2.3) and (2.4) with
\begin{equation}
\lambda_i\equiv K_i (d_i-1).
\end{equation}
For the metric (2.1) the Ricci scalar curvature of $M$ is
\begin{equation}
R=e^{-2\gamma}\left \{
\left [ \sum_{i=1}^{n} (d_i \dot\beta^i) \right ]^2
+ \sum_{i=1}^{n}
d_i[ {(\dot\beta^i)^2-2\dot\gamma\dot\beta^i+2\ddot \beta^i} ]
\right \}
+\sum_{i=1}^{n} R^{(i)} e^{-2\beta^i}.
\end{equation}

Let us now consider a variation principle with the action
\begin{equation}
S=S_{EH}+S_{GH}+S_{M}
\end{equation}
where $S_M$ is a matter term,
\[
S_{EH}=\frac{1}{2\kappa^2}\int_{M}\sqrt{\vert g\vert} R\, dx
\]
is the Einstein-Hilbert action and
\[
S_{GH}=\frac{1}{\kappa^2}\int_{\partial M}\sqrt{\vert h\vert} K \, dy
\]
is the Gibbons-Hawking boundary term$^\Gi$, where $K$ is the trace
of the second fundamental form, which just cancels
second time derivatives in the equation of motion.

Let us define a metric on the minisuperspace, which is spanned here in
the coordinates $\beta^i$. We set
\begin{equation}
G_{ij}:=d_i \delta_{ij}-d_i d_j,
\end{equation}
thus defining the components
$G_{ij}$ of the minisuperspace metric
\begin{equation}
G = G_{ij}d\beta^i\otimes d\beta^j.
\end{equation}
Furthermore we define
the minisuperspace lapse function by
\begin{equation}
N:=e^{\gamma-\sum_{i=1}^{n}d_i\beta^i}
\end{equation}
and a minisuperspace potential $V=V(\beta^i)$ via
\begin{equation}
V:=-\frac{\mu}{2}\sum_{i=1}^{n}R^{(i)}
e^{-2\beta^i+\gamma+\sum_{j=1}^{n}d_j\beta^j},
\end{equation}
where
\begin{equation}
\mu:=\kappa^{-2}\prod_{i=1}^{n}\sqrt{\vert\det g^{(i)}\vert}.
\end{equation}
Then the
variational principle of (2.10) is  equivalent to
a Lagrangian variational principle in minisuperspace,
\begin{equation}
S=\int Ldt, \quad{\rm where}\quad
L=N\{\frac{\mu}{2}N^{-2}G_{ij}\dot\beta^i\dot\beta^j-V\}.
\end{equation}
Here $\mu$ is the mass of a classical particle in minisuperspace.
Note that $\mu^2$ is proportional to the volumes of spaces $M_i$.

Next let us compare different choices of time $\tau$  in Eq. (2.1).
The time gauge is determined by the function $\gamma$.
There exist few natural time gauges from the physical point of view.

i) The {\em synchronous time gauge}
\begin{equation}
\gamma\equiv 0,
\end{equation}
for which $t$ in Eq. (2.1) is the proper time $t_s$ of the universe.
The clocks of geodesically comoved observers go synchronous to that
time.

ii) The {\em conformal time gauges} on $\R\times M_i\subset M$
\begin{equation}
\gamma\equiv \beta^i,
\end{equation}
for which $t$ in Eq. (2.1) is the conformal time $\eta_i$ of $M_i$
for some $i\in \{1,\ldots,n\}$,
given by
\begin{equation}
d\eta_i=e^{-\beta^i}dt_s.
\end{equation}

iii) The {\em mean conformal time gauge} on $M$:
\nl
For $n>1$ and $\beta^2\neq\beta^1$
on $M$ the usual concept of a conformal time does no longer apply.
Looking for a generalized ``conformal time" $\eta$ on $M$, we set
\begin{equation}
d:=D-1=\sum_{i=1}^n d_i
\end{equation}
and consider the gauge
\begin{equation}
\gamma\equiv \frac{1}{d}\sum_{i=1}^n d_i\beta^i,
\end{equation}
which yields a time $t\equiv\eta$ given by
\begin{equation}
d\eta=\left( \prod_{i=1}^n a_i^{d_i} \right)^{-1/d} dt_s.
\end{equation}
Here $\prod_{i=1}^n a_i^{d_i}$ is proportional to the volume of
$d$-dimensional spacial sections in $M$
and the relative time scale factor
\begin{equation}
\left( \prod_{i=1}^n a_i^{d_i} \right)^{1/d}
=e^{\frac{1}{d}\sum_{i}d_i\beta^i}
\end{equation}
is given by a scale exponent, which is the dimensionally weighted arithmetic
mean of the spacial scaling exponents of spaces $M_i$.
It is
\begin{equation}
(dt_s)^d= e^{\sum_{i}d_i\beta^i} d\eta^d.
\end{equation}
Since on the other hand by Eq. (2.19) we have
\begin{equation}
(dt_s)^d=\otimes_{i=1}^n \left( e^{\beta^i}d\eta_i \right)^{d_i},
\end{equation}
together with Eq. (2.24) we yield
\begin{equation}
(d\eta)^d=e^{-\sum_{i}d_i\beta^i}
\otimes_{i=1}^n \left( e^{\beta^i}d\eta_i \right)^{d_i}.
\end{equation}
So the time $\eta$ is a mean conformal time, given differentially
as a dimensionally scale factor weighted geometrical tensor
average of the conformal times $\eta_i$.
An alternative to the mean conformal time $\eta$
is given by a similar differential averaging like Eq. (2.26), but weighted
by an additional factor of $e^{(1-d)\sum_{i}d_i\beta^i}$.
This is gauge is described in the following.

iv) The {\em harmonic time gauge}
\begin{equation}
\gamma\equiv\gamma_h:=\sum_{i=1}^n d_i\beta^i
\end{equation}
yields the time $t\equiv t_h$, given by
\begin{equation}
dt_h=\left( \prod_{i=1}^n a_i^{d_i} \right)^{-1} dt_s
=\left( \prod_{i=1}^n a_i^{d_i} \right)^{\frac{1-d}{d}} d\eta.
\end{equation}
In this gauge any function $\varphi$ with $\varphi (t,y)=t$
is harmonic, i.e. $\Delta[g] \varphi =0$, and
the minisuperspace lapse function is $N\equiv 1$.
The latter is especially convenient when we work in
minisuperspace.

Here and in the following for any $x$ we set
\begin{equation}
\dot x:=\frac{\partial x}{\partial t_h},
\end{equation}
so the dot denotes the partial derivative w.r.t. harmonic time.

Then the equations of motion from Eq. (2.16) yield
\begin{equation}
\mu G_{ij}\ddot\beta^j=-\frac{\partial V}{\partial\beta^i}
\end{equation}
plus the energy constraint
\begin{equation}
\frac{\mu}{2}G_{ij}\dot\beta^i\dot\beta^j+V=0.
\end{equation}

Now we consider the Einstein equations for a universe (2.1),
given by
\begin{equation}
R_{\mu\nu}-\frac{1}{2}R g_{\mu\nu}=T_{\mu\nu}
\end{equation}
with energy momentum tensor $T_{\mu\nu}$ corresponding to $S_M$.
Let us assume we that the energy momentum tensor $T_{\mu\nu}$
is of perfect fluid type, depending only on the matter density $\rho$
in the universe and on the
pressures $p_i$ in the spaces $M_i$.
With equations of state
\begin{equation}
p_i=p_i(\rho)
\end{equation}
for the pressures $p_i$ in $M_i$ in terms of the matter density
\begin{equation}
\rho=\rho(\beta^i),
\end{equation}
the energy momentum tensor is a function of the dynamical variables
$\beta^i$.

The continuity equation
$\frac{\partial \rho}{\partial t}=d_j\dot\beta^j(\rho+p_j)$
together with an equation of state
\begin{equation}
p_j=(\frac{m_j}{d_j}-1)\rho
\end{equation}
yields
\begin{equation}
\rho=\rho(\beta^i)=M_{m_1\cdots m_n}e^{-\sum_{j=1}^{n}m_j\beta^j}.
\end{equation}
For tracefree $T_{\mu\nu}$ we have
\begin{equation}
\sum_{j=1}^{n} m_j=d=D-1.
\end{equation}
Let us consider the Ricci scalar curvatures $R^{(i)}$ of $M_i$ as the
only sources of (stress-)energy.
Then the density is
\begin{equation}
\rho=-\sum_{i=1}^{n}R^{(i)}e^{-2\beta^i},
\end{equation}
which is positive resp. negative, if all $R^{(i)}$ are negative
resp. positive semidefinite and at least $R^{(1)}\neq 0$.
With such a density $\rho(\beta^i)$ the minisuperspace potential
of Eqs. (2.14) can be written as
\begin{equation}
V=\frac{\mu}{2}\rho(\beta^i)e^{\gamma+\sum_{i=1}^{n}d_i\beta^i}.
\end{equation}

The equations of motion are known to be integrable
if spaces $M_i$ are flat for all $i>1$.
Therefore in the following we restrict to models with $R^{(i)}=0$ for
$i>1$.

Up to now we have considered by (2.34) only a dependence of $\rho$ on
the the geometrical data given by the $\beta^i$ in Eq. (2.1). More
generally we will admit in the following also a dependence on a
scalar field $\Phi$.
This field shall be minimally coupled to the geometry of minisuperspace
and have a potential $U(\Phi)$.

The Lagrangian variational principle is
given in this case by a Lagrangian
$$
L=\frac{1}{2}{\mu}
e^{-\gamma+\sum_{i=1}^{n}d_i\beta^i}
\left \{
\sum_{i=1}^{n}{d_i(\dot\beta^i)^2}
-[\sum_{i=1}^{n}{d_i\dot\beta^i}]^2
+\kappa^2{\dot \Phi}^2
\right \}
$$
\begin{equation}
+\frac{1}{2}{\mu}
e^{\gamma+\sum_{i=1}^{n}d_i\beta^i}R^{(1)}e^{-2\beta^1}
-{\mu}\kappa^2 e^{\gamma+\sum_{i=1}^{n}d_i\beta^i}
U(\Phi),
\end{equation}
where by Eq. (2.15) the mass
$\mu=\kappa^{-2}\prod_{i=1}^{n}\sqrt{\vert\det g^{(i)}\vert}$
in minisuperspace is actually determined by the volumes of the spaces $M_i$.

\section{\bf Conformal Lagrangian Models and their Solutions}
\setcounter{equation}{0}
Generally we will have to distinguish between (1) conformal
transformations of {\em Lagrangian models} and
(2) transformations of the solutions of a fixed given Lagrangian model to
a conformal {\em coordinate frame}.

(1) Conformal transformations of Lagrangian models:
\nl
We consider a differentiable manifold $M$. Equipped with a Riemannian
structure $g_{ij}$ and scalar fields $(\phi^1,\ldots,\phi^k)$ on $M$
we obtain a {\em Lagrangian model} by imposing a Lagrangian variation
principle
\begin{equation}
\delta S=0 \quad{\rm with}\quad S=\int_{M}  \sqrt{\vert g\vert}{L} d^D\!x
\end{equation}
given by a second order Lagrangian
\begin{equation}
L=L(g_{ij},\phi^1,\ldots,\phi^k;
    g_{ij,l},\phi^1_{,l},\ldots,\phi^k_{,l};
    g_{ij,lm}).
\end{equation}
Conformal transformation {\em of the model} keeps $M$ fixed as a
differentiable manifold,
but varies its additional structures conformally
\begin{equation}
(g_{ij},\phi^1,\ldots,\phi^k)\to (\hat g_{ij},\hat\phi^1,\ldots,\hat\phi^k),
\end{equation}
yielding a new variational principle by demanding
\begin{equation}
\sqrt{\vert g\vert}{L} \doteq \sqrt{\vert \hat g\vert}{\hat L}
\end{equation}
for the new Lagrangian
\begin{equation}
\hat L=\hat L(\hat g_{ij},\hat\phi^1,\ldots,\hat\phi^k;
\hat g_{ij,l},\hat\phi^1_{,l},\ldots,\hat\phi^k_{,l};
\hat g_{ij,lm})
\end{equation}
Therefore conformal transformation of models are performed in practice
on a {\em fixed coordinate patch} $x^i$ of $M$.

(2) Conformal transformation of solutions to new coordinates:
\nl
We {\em fix a Lagrangian model} and transform the metric tensor components
conformally,
\begin{equation}
g_{i'j'}=e^{2f(x)}g_{kl},
\end{equation}
via a coordinate transform satisfying
\begin{equation}
dx'^{i}=e^{-f(x)}dx^i \quad{\rm or}\quad
\frac{\partial x'^{i}}{\partial x^j}=e^{-f(x)}\delta^i_j.
\end{equation}
Here the first fundamental form
\begin{equation}
ds^2=g_{i'j'}dx^{i'}dx^{j'}=g_{ij}dx^{i}dx^{j},
\end{equation}
and therefore the inner geometry, remains invariant, in contrast to
transformations under (1) above.
Since all geometric invariants remain unchanged, the model is still the
same, though looking different in  different coordinate frames.

A special application of transformations (2) are time gauge
transformations from arbitrarily given coordinates to
one of the natural time gauges (i) to (iv).
Via transformations (2) for any universe (2.1) e.g. there exists a frame
which is
w.r.t. time either synchronous (2.17) or harmonic (2.27), though in practice
this frames may be difficult to compute explicitly. We will come back to
this point
later.

Now we want to study the effect of transformations (1) in more detail.
One application of special interest is the transformation from
a Lagrangian model with minimally coupled potential to a conformally
equivalent one with nonminimal coupled potential and vice versa.

Let us follow Ref. \Ma\  and consider an action of the kind
\begin{equation}
S=\int d^Dx\sqrt{\vert g\vert}(F(\phi,R)-\frac{\epsilon}{2}(\nabla\phi)^2).
\end{equation}
With
\begin{equation}
\omega:=\frac{1}{D-2}\ln(2\kappa^2
			  \vert\frac{\partial F}{\partial R}\vert)+C
\end{equation}
the conformal factor
\begin{equation}
e^{\omega}=
[2\kappa^2\vert\frac{\partial F}{\partial R}\vert]^\frac{1}{D-2}e^{C}
\end{equation}
yields a conformal transformation from $g_{\mu\nu}$
to the minimal metric
\begin{equation}
\hat g_{\mu\nu}=e^{2\omega}g_{\mu\nu}.
\end{equation}

Especially let us consider in the following actions, which are linear
in $R$. With
\begin{equation}
F(\phi,R)=f(\phi)R-V(\phi).
\end{equation}
the action is
\begin{equation}
S=\int d^Dx\sqrt{\vert g\vert}(f(\phi)R-V(\phi)
-\frac{\epsilon}{2}(\nabla\phi)^2).
\end{equation}

The minimal metric is then related to the conformal one by (3.12) with
\begin{equation}
\omega=\frac{1}{D-2}\ln(2\kappa^2
\vert f(\phi)\vert)+C
\end{equation}
The scalar field in the minimal model is
$$
\Phi=\kappa^{-1}\int d\phi\{\frac{\epsilon(D-2)f(\phi)+2(D-1)(f'(\phi))^2}
			       {2(D-2)f^2(\phi)}\}^{1/2}  =
$$
\begin{equation}
=(2\kappa)^{-1}\int d\phi\{\frac{2\epsilon f(\phi)+\xi_c^{-1}(f'(\phi))^2}
			       {f^2(\phi)}\}^{1/2},
\end{equation}
where
\begin{equation}
\xi_c:=\frac{D-2}{4(D-1)}
\end{equation}
is the conformal coupling constant.

For the following we define $\sign x$ to be $\pm 1$ for $x\geq 0$ resp. $x<0$.
Then with the new minimally coupled potential
\begin{equation}
U(\Phi)=(\sign f(\phi))\ [2\kappa^2\vert f(\phi)\vert]^{-D/D-2}V(\phi)
\end{equation}
the corresponding minimal action is
\begin{equation}
S=\sign f\int d^Dx\sqrt{\vert \hat{g}\vert}\left(-\frac{1}{2}
	  [(\hat{\nabla}\Phi)^2-\frac{1}{\kappa^2}\hat{R}]-U(\Phi)\right).
\end{equation}

Example 1:
\begin{equation}
f(\phi)=\frac{1}{2}\xi\phi^2,
\end{equation}
\begin{equation}
V(\phi)=-\lambda\phi^\frac{2D}{D-2}.
\end{equation}
Substituting this into Eq. (3.18) the corresponding minimal potential
$U$ is constant,
\begin{equation}
U(\Phi)=(\sign \xi)\ \vert\xi\kappa^2\vert^{-D/D-2} \,\lambda.
\end{equation}
It becomes zero precisely for $\lambda=0$, i.e. when $V$ is zero.
With
\begin{equation}
f'(\phi)=\xi\phi
\end{equation}
we obtain
$$
\Phi=\kappa^{-1}\int d\phi
\left\{
\frac{ ({\epsilon\over\xi} + {1\over{\xi_c}} ) \phi^2}{\phi^4}
\right\}^\frac{1}{2}
=\left(\kappa\sqrt\xi\right)^{-1}
\sqrt{\frac{1}{\xi_c}+ {\epsilon\over\xi} }
\int d\phi\frac{1}{\vert\phi\vert}
$$
\begin{equation}
=\kappa^{-1}\sqrt{\frac{1}{\xi_c}+{\epsilon\over\xi} }\,\ln\vert\phi\vert
+C
\end{equation}
for $-\frac{\xi}{\epsilon}\geq\xi_c$.
Note that for
\begin{equation}
\frac{\xi}{\epsilon}=-\xi_c,
\end{equation}
e.g. for $\epsilon=-1$ and conformal coupling, we have
\begin{equation}
\Phi=C.
\end{equation}
Thus here the conformal coupling theory is equivalent to
a theory without scalarfield.

For $-\frac{\xi}{\epsilon}<\xi_c$ the field $\Phi$ would become
complex and, for imaginary $C$, purely imaginary.

In any case the integration constant $C$ may be a function of the
coupling $\xi$ and the dimension $D$.

Example 2:\nl
\begin{equation}
f(\phi)=\frac{1}{2}(1-\xi\phi^2),
\end{equation}
\begin{equation}
V(\phi)=\Lambda.
\end{equation}
Then the constant potential $V$ has its minimal correspondence in a
non constant $U$, given by
\begin{equation}
U(\Phi)=\pm \Lambda \vert\kappa^2 (1-\xi\phi^2) \vert^{-D/D-2}
\end{equation}
respectively
for $\phi^2<\xi^{-1}$ or $\phi^2>\xi^{-1}$.

Let us set in the following
\begin{equation}
\epsilon=1.
\end{equation}
Then with
\begin{equation}
f'(\phi)=-\xi\phi
\end{equation}
we obtain
\begin{equation}
\Phi=\kappa^{-1}\int d\phi\{\frac{1+c\,\xi\phi^2}
			       {(1-\xi\phi^2)^2}\}^{1/2},
\end{equation}
where
\begin{equation}
c:=\frac{\xi}{\xi_c}-1.
\end{equation}

For $\xi=0$ it is $\Phi=\kappa^{-1}\phi +A$, i.e. the coupling remains
minimal.

To solved this integral for $\xi\neq 0$, we substitute $u:=\xi\phi^2$.

To assure a solution of (3.32) to be real, let us assume $\xi\geq\xi_c$
which yields $c\geq 0$.

Then we obtain
$$
\Phi=\frac{\sign(\phi)}{2\kappa\sqrt{\xi}}
		  \int{\frac {\sqrt {u^{-1}+c}}{\vert1-u\vert}}du
+C_{<\atop>}
$$
$$
=\frac{\sign((1-u)\phi)}{2\kappa\sqrt{\xi}}
[-\sqrt {c}\ln (2\,\sqrt {c}\sqrt {1+cu}\sqrt {u}+2\,cu+1)+
$$
$$
\sqrt {1+c}
\ln ({\frac{2\,\sqrt{1+c}\sqrt{1+cu}\sqrt {u}+2\,cu+1+u}{\vert1-u\vert}})]
+C_{<\atop>}
$$
$$
=\frac{\sign((1-\xi\phi^2)\phi)}{2\kappa\sqrt{\xi}}
\{-\sqrt {c}\ln (2\,\sqrt {c}\sqrt {1+c\xi\,\phi^{2}}
\sqrt {\xi} \vert\phi\vert
+2\,c\xi\,\phi^{2}+1)
$$
$$
+ \sqrt {1+c}\ln ({\frac {2\,\sqrt {1+c}
\sqrt{1+c\xi\,\phi^{2}} \sqrt {\xi} \vert\phi\vert
+2\,c\xi\,\phi^{2}+1+\xi\,\phi^{2}}
{\vert 1-\xi\,\phi^{2}\vert}})\}
+C_{<\atop>}
$$
$$
=\frac{\sign((1-\xi\phi^2)\phi)}{2\kappa\sqrt{\xi}}
\ln
\frac{ [2\,\sqrt {1+c} \sqrt{1+c\xi\,\phi^{2}}\sqrt {\xi}\vert\phi\vert
	    +(2\,c+1)\xi\,\phi^{2}+1]^{\sqrt {1+c}}  }
{ [2\,\sqrt {c}\sqrt {1+c\xi\,\phi^{2}}\sqrt {\xi}\vert\phi\vert
	     +2\,c\xi\,\phi^{2}+1]^{\sqrt{c}}
\cdot {\vert 1-\xi\,\phi^{2}\vert}^{\sqrt {1+c}}  }
$$
\begin{equation}
+C_{<\atop>}.
\end{equation}
The integration constants $C_{<\atop>}$
for $\phi^2<\xi^{-1}$ and $\phi^2>\xi^{-1}$ respectively
may be arbitrary functions of $\xi$ and the dimension $D$.

The singularities of the transform $\phi\to\Phi$ are located at
$\phi^2=\xi^{-1}$.

If the coupling is conformal $\xi=\xi_c$,
i.e. $c=0$, the expressions (3.34) simplify to
\begin{equation}
\kappa\Phi=\frac{1}{\sqrt{\xi_c}} [(\artanh\sqrt{\xi_c}\phi)+c_<]
\end{equation}
for $\phi^2<\xi^{-1}_c$ and to
\begin{equation}
\kappa\Phi=\frac{1}{\sqrt{\xi_c}} [(\arcoth\sqrt{\xi_c}\phi)+c_>]
\end{equation}
for $\phi^2>\xi^{-1}_c$.

In the following we restrict to this case of conformal coupling.

The inverse formulas expressing the conformal field $\phi$ in terms of
the minimal field $\Phi$ are
\begin{equation}
\phi=\frac{1}{\sqrt{\xi_c}} \left[ \tanh(\sqrt{\xi_c}\kappa\Phi-c_<) \right]
\end{equation}
with $\phi^2<\xi^{-1}_c$ and
\begin{equation}
\phi=\frac{1}{\sqrt{\xi_c}} \left[ (\coth(\sqrt{\xi_c}\kappa\Phi-c_>) \right]
\end{equation}
with $\phi^2>\xi^{-1}_c$ respectively.

The conformal factor is according to Eqs. (3.15) and (3.27) given by
\begin{equation}
\omega=\frac{1}{D-2}\ln(\kappa^2 \vert 1-\xi_c\phi^2 \vert)+C.
\end{equation}

In the following we want to compare the solutions of the minimal model
to those of the corresponding conformal model.
We specify the geometry for the minimal model to be of multidimensional
type (2.1), with all $M_i$ Ricci flat, hence $R^{(i)}=0$ for $i=1,\ldots,n$.
The minimally coupled scalar field is assumed to have zero potential
$U\equiv 0$.
In the harmonic time gauge (2.27) with harmonic time
\begin{equation}
\tau\equiv t^{(m)}_h,
\end{equation}
we demand this model to be a solution for Eq. (2.40)
with vanishing $R^{(1)}$ and $U(\Phi)$.
We set $\beta^{n+1}:= \kappa \Phi$ and
and obtain as solution
a multidimensional (Kasner like) universe,
given by
\begin{equation}
\hat\beta^i=b^i\tau+c^i  \ \mbox{and}\
\hat\gamma=\sum_{i=1}^n d_i \hat\beta^i
=(\sum_{i=1}^n d_i b^i)\tau+(\sum_{i=1}^n d_i c^i),
\end{equation}
with $i=1,\ldots,n+1$,
where with $V\equiv 0$ the constraint Eq. (2.31) simply
reads
\begin{equation}
G_{ij} b^i b^j + (b^{n+1})^2=0.
\end{equation}

With Eq. (3.39) the scaling powers of the universe
given by Eqs. (3.41) with $i=1,\ldots,n$ transform to corresponding
scale factors of the conformal universe
$$
\beta^i=\hat\beta^i-\omega
$$
\begin{equation}
=b^i\tau+\frac{1}{2-D}\ln\vert 1-\xi_c(\phi)^2\vert
+ c^i + \frac{2}{2-D} \ln\kappa-C
\end{equation}
and
$$
\gamma=\sum_{i=1}^n d_i \beta^i
$$
\begin{equation}
=(\sum_i d_i b^i)\tau+\frac{1}{2-D}\ln\vert 1-\xi_c(\phi)^2\vert
+(\sum_i d_i c^i) + \frac{2}{2-D} \ln\kappa-C.
\end{equation}
It should be clear from the remarks in Sec. 3 that
the variable $\tau$, when harmonic in the minimal model,
in the conformal model cannot be expected to be harmonic either,
i.e. in general
\begin{equation}
\tau\neq t^{(c)}_h.
\end{equation}
Let us take for simplicity
\begin{equation}
C=\frac{2}{2-D} \ln\kappa,
\end{equation}
which yields the lapse function
\begin{equation}
e^\gamma=e^{(\sum_i d_i b^i)\tau+(\sum_i d_i c^i)}
\vert 1-\xi_c(\phi)^2\vert^{\frac{1}{2-D}}
\end{equation}
and for $i=1,\ldots,n$ the scale factors
\begin{equation}
e^{\beta^i}= e^{b^i\tau+ c^i}
\vert 1-\xi_c(\phi)^2\vert^{\frac{1}{2-D}}.
\end{equation}
Let us further set for simplicity
\begin{equation}
c_{<}=c_{>}=\sqrt{\xi_c} c^{n+1}.
\end{equation}

The transformation of the scalar field from the
solution (3.39) of the minimally coupled model
\begin{equation}
\kappa\Phi(\tau)=b^{n+1}\tau+c^{n+1}
\end{equation}
to the scalarfield of the conformal model by Eqs. (3.37) or (3.38)
and substitution of the latter in Eqs. (3.48) resp. (3.49) yields
a lapse function
\begin{equation}
e^\gamma=e^{(\sum_i d_i b^i)\tau+(\sum_i d_i c^i)}
\cosh^{\frac{2}{D-2}}( \sqrt{\xi_c} b^{n+1}\tau )
\end{equation}
resp.
\begin{equation}
e^\gamma=e^{(\sum_i d_i b^i)\tau+(\sum_i d_i c^i)}
\vert\sinh^{\frac{2}{D-2}}( \sqrt{\xi_c} b^{n+1}\tau )\vert
\end{equation}
and, with $i=1,\ldots,n$, nonsingular scale factors
\begin{equation}
e^{\beta^i}= e^{b^i\tau+ c^i}
\cosh^{\frac{2}{D-2}}( \sqrt{\xi_c} b^{n+1}\tau )
\end{equation}
resp. singular scale factors
\begin{equation}
e^{\beta^i}= e^{b^i\tau+ c^i}
\vert\sinh^{\frac{2}{D-2}}( \sqrt{\xi_c} b^{n+1}\tau )\vert
\end{equation}
of the conformal model. The scale factor singularity
of the minimal coupling model for $\tau\to-\infty$ vanishes in the conformal
model of Eqs. (3.51) and (3.53) for a scalar field $\phi$ bounded
according to (3.37).
For $D=4$ this result had already been indicated by Ref. \Ga.

On the other hand in the conformal model
of Eqs. (3.52) and (3.54), with $\phi$ according to (3.38),
though the scale factor singularity of the minimal model
for $\tau \to -\infty$ has also disappeared, instead there is another new
scale factor singularity at finite
(harmonic) time $\tau=0$.

Let us consider a special case of the nonsingular solution
with $\phi^2<\xi_c^{-1}$,
where we assume the internal spaces
to be static in the minimal model, i.e.  $b^i=0$ for $i=2,\ldots,n$.
Then in the conformal model, the internal spaces are no longer
static. Their scale factors (3.54) with $i>2$ have a minimum at
$\tau=0$.
Remind that for solution (3.41) all spaces $M_i$,
internal and external, $i=1,\ldots,n$ have been assumed as flat.
{}From Eq. (3.42) with $G_{11}=d_1(1-d_1)$ we find
that the scalar field is given by
\begin{equation}
(b^{n+1})^2= d_1 (d_1-1) (b^1)^2.
\end{equation}

With real $b_1$ then also
\begin{equation}
b^{n+1}=\pm\sqrt{d_1 (d_1-1)}b^1
\end{equation}
is real and by Eq. (3.52) the scale $a_1$ of $M_1$ has a minimum at
\begin{equation}
\tau_0
=(\sqrt{\xi_c} b^{n+1})^{-1}\artanh\left(
\frac{(2-D)}{2\sqrt{\xi_c}}
\frac{b^1}{b^{n+1}} \right),
\end{equation}
with $\tau_0>0$ for $b^1<0$ and $\tau_0<0$ for $b^1>0$.

The points $\tau=\tau_0$ and $\tau=0$ are the turning points
in the minimum for the factor spaces $M_1$ and $M_2,\ldots,M_n$
respectively.
It is interesting to explain the creation of our Lorentzian universe
by a ''birth from nothing''$^\Vi$, i.e. quantum tunneling from an Euclidean
region.
Let us first consider the geometry of this tunneling as usual for the
external universe $\R\times M_1$.
So if we cut $M$
along the minimal hypersurface at $\tau_0$ in 2 pieces, one of them,
say $M'$, contains
the hypersurface $\tau=0$ where the internal spaces are minimal.
We set $M'':=M\setminus M'$ to be the remaining piece.
Then we can choose (eventually with time reversal $\tau\to-\tau$)
either $M'$ or $M''$ as a universe $\tilde M$ that is generated at $\tau_0$
with initial minimal scale $a_1(\tau_0)$.
In the usual quantum tunneling interpretation, at the
scale $a_1(\tau_0)$ with $\dot a_1(\tau_0)=0$
one glues smoothly a compact simply connected Euclidean space-time region to
the Lorentzian $\tilde M$, yielding a joint differentiable manifold $\hat M$.
Then the sum of classical paths passing the boundary $\partial\tilde M$
from the Euclidean to the Lorentzian region
can be interpreted as quantum tunneling from ''nothing''$^\Vi$ to $\tilde M$.

According to
Ref. $\aRa$ this interpretation has a direct topological
correspondence in a projective blow up of a singularity of shape
$M_2\cdots \times M_n$ (the ''nothing'') to
$S^{d_1}(a_1(\tau_0))\times M_2\cdots \times M_n$, where
$S^{d_1}(a_1(\tau_0))$ denotes the $d_1$-dimensional sphere of radius
$a_1(\tau_0)$.

For $\tilde M=M'$ the internal spaces shrink for (harmonic) time
from $\tau_0$ towards $\tau=0$ and expand from $\tau=0$ onwards for ever,
but for $\tilde M=M''$ the internal spaces expand for (harmonic) time
from $\tau_0$ onwards for ever.
So the decomposition of $M$ in $M'$ and $M''$ is highly asymmetric
w.r.t. the internal spaces. For more realistic
models it might be especially useful to consider
the piece of $M'$ which lies between $\tau_0$ and $\tau=0$,
since it can describe a shrinking of internal spaces while the
external space is expanding.

Remarkably the multidimensional geometries
with $\tau<\tau_0$ and $\tau>\tau_0$ are $\tau$-asymmetric to each
other. Taking one as contracting, the other as expanding
w.r.t. $M_1$, the two are distinguished
by a qualitatively different behavior of internal spaces $M_k$, $k\geq 2$.

The latter allows to choose the ''arrow of time''$^\Ze$ in a natural
manner determined by  intrinsic features of the solutions.
Note if there is at least one
internal extra space, i.e. $n>1$, then the minisuperspace w.r.t.
scalefactors of geometry
has Lorentzian signature $(-,+,\ldots,+)$. After diagonalization
of (2.11) by a minisuperspace coordinate transformation
$\beta^i\to\alpha^i$ ($i=1,\ldots,n$), there is just one new scalefactor
coordinate, say $\alpha^1$, which corresponds to the
negative eigenvalue of $G$, and hence
assumes the role played by time in usual quantum mechanics.
(For $n=1$ there are no internal spaces, but $G_{11}<0$ for
$d_1>1$ still provides a negative eigenvalue that is distinguished
at least against the additional positive eigenvalue from the scalar
field.) This shows that, at least after diagonalization,
that an ''external'' space
is distinguished against the internal spaces, because its scale factor
provides a natural ''time'' coordinate.

Upto now we have considered the smooth tunneling from
an Euclidean region to the external universe $\R\times M_1$,
where the external spaces have been considered as purely
passive spectators of the tunneling process.
As we have pointed out in contrast to models with only one (external) space
factor $M_1$, the additional internal spaces $M_2,\ldots,M_n$
yield an asymmetry of $M$ w.r.t. (harmonic) time $\tau$ for
$\tau_0\neq 0$, which is according to Eq. (3.57) the case exactly when
$D\neq 2$ and the external space is non static, i.e. $b_1\neq 0$.

In the following we want to obtain a quantum tunneling interpretation
for all of $M$, including the internal spaces. The picture
becomes more complicated, since the extremal hypersurfaces of
external space and internal spaces are located at different times
$\tau=\tau_0$ resp. $\tau=0$.

Let $M_1$ be the external space with $b_1>0$ and hence $\tau_0<0$.
Let us start with an Euclidean region of complex geometry
given by scale factors
$$
a_k=e^{-ib^k\tau+\tilde c^k}
\vert\sin ( \sqrt{\xi_c} b^{n+1}\tau )\vert^{\frac{2}{D-2}}.
$$
Then we can perform an analytic continuation to the Lorentzian region
with $\tau\to i\tau+\pi/(2 \sqrt{\xi_c} b^{n+1})$, and we require
$c^k=\tilde c^k-i\pi b^k/(2 \sqrt{\xi_c} b^{n+1})$ to be the real
constant of the real geometry (3.48).

The quantum creation (via tunneling)
of different factor spaces takes place at different values of $\tau$
(see Fig. 1).

%

First the factor space $M_1$ comes into real existence and after an
time interval $\Delta\tau=\vert\tau_0\vert$ the internal factor
spaces $M_2, \ldots, M_n$ appear in the Lorentzian region. Since
$\Delta\tau$ is arbitrarily large, there is in principle an alternative
explanation of the unobservable extra dimensions, independent
from concepts of compactification and shrinking to a fundamental length
in symmetry breaking. Similar to the spirit of the
idea that internal dimensions might be hidden
due to a potential barrier$^\RS$, they may have been up to now still
in the Euclidean region and hence unobservable. This view is also
compatible with their interpretation as complex resolutions of
ADE symmetries$^\aRa$.

Now let us perform a transition from Lorentzian time $\tau$ to Euclidean
time $i\tau$. Then with a simultaneous transition from $b^k$ to
$- i b^k$
for $k=1,\ldots,n$ the geometry remains real, since
$\hat\beta^k=b^k\tau+c^k$ is unchanged.
But the analogue of Eq. (3.56) for the Euclidean region
then becomes
\begin{equation}
b^{n+1}=\mp i\sqrt{d_1 (d_1-1)}b^1.
\end{equation}
Hence the scalar field is purely imaginary.
This solution corresponds to a classical (instanton) wormhole.
The sizes of the wormhole throats in the factor spaces $M_2,\ldots,M_n$
coincide with the sizes of static spaces in the
minimal model, i.e. $\hat a_2(0),\ldots,\hat a_n(0)$ respectively.

With Eq. (3.56) replaced by (3.58), the Eq. (3.57)
remains unchanged in the transition to the Euclidean region,
and the minimum of the scale $a_1$ (unchanged geometry !)
now corresponds to the throat of the wormhole.

If one wants to compare the synchronous time pictures of the
minimal and the conformal msolution, one has to calculate them for
both metrics.
In the minimal model
we have
\begin{equation}
dt^{(m)}_s=e^{\hat\gamma}d\tau
=e^{(\sum_i d_i b^i)\tau+(\sum_i d_i c^i)}d\tau,
\end{equation}
which can be integrated to
\begin{equation}
t^{(m)}_s=(\sum_i d_i b^i)^{-1}
e^{\hat\gamma}+t_0.
\end{equation}

The latter can be inverted to
\begin{equation}
\tau=(\sum_i d_i b^i)^{-1}
\left\{[\ln(\sum_i d_i b^i)(t^{(m)}_s-t_0)]-(\sum_i d_i c^i)\right\}.
\end{equation}
Setting
\begin{equation}
B:=\sum_{i=1}^{n} d_i b^i \ \mbox{and} \ C:=\sum_{i=1}^{n} d_i c^i,
\end{equation}
this yields the scale factors
\begin{equation}
\hat{a_s}^i=(t^{(m)}_s-t_0)^{b^i/B} e^{\frac{b_i}{B}(\ln B-C)+c_i}
\end{equation}
and the scalarfield
\begin{equation}
\kappa\Phi=\frac{b^{n+1}}{B} \{ [\ln B(t^{(m)}_s-t_0)] -C \}+c^{n+1}.
\end{equation}
Let us define for $i=1,\ldots,n+1$ the numbers
\begin{equation}
\alpha^i:=\frac{b^i}{B}.
\end{equation}
With (3.62) they satisfy
\begin{equation}
\sum_{i=1}^{n} d_i \alpha^i=1,
\end{equation}
and by Eq. (3.42) also
\begin{equation}
\alpha^{n+1}=\sqrt{1-\sum_{i=1}^{n} d_i (\alpha^i)^2}.
\end{equation}
Eqs. (3.63) shows, that the solution
(3.41) is really a generalized Kasner universe with exponents $\alpha^i$
satisfying generalized Kasner conditions (3.66) and (3.67).

In the conformal model the synchronous time is given as
\begin{equation}
t^{(c)}_s=\int e^\gamma d\tau=
\int\cosh^{\frac{2}{D-2}}(\sqrt{\xi_c} b^{n+1}\tau)
e^{B\tau+C} d\tau
\end{equation}
resp.
\begin{equation}
t^{(c)}_s=\int e^\gamma d\tau=
\int \sinh^{\frac{2}{D-2}}(\sqrt{\xi_c} b^{n+1}\tau)
e^{B\tau+C} d\tau.
\end{equation}
Similarily one could also try to calculate other time gauges for
both metrics.

\section{\bf Quantum Solutions from the WdW equation}
\setcounter{equation}{0}
In this section we investigate the quantum analogue of the classical
solution for the particular model of Sec. 3 above with all
$M_i$ Ricci flat.
The WdW equation for the minimal model reeds$^\IMZ$
\begin{equation}
\left[ G^{ij}\frac{\partial}{\partial\hat\beta^i}
      \frac{\partial}{\partial\hat\beta^j}
      +\frac{\partial^2}{\partial\Phi^2} \right]\, \Psi=0,
\end{equation}
where the minimally coupled field $\Phi$ is redefined by
$\Phi:\gets\kappa \Phi$ as compared with the previous section,
its potential is zero, $U(\Phi)\equiv 0$, and the WdW equation
is written in harmonic time gauge$^\bRa$, with  components
$ G^{ij}=\frac{\delta_{ij}}{d_i}-\frac{1}{2-D}$ of the inverse
to the minisuperspace metric (2.11-12)
in coordinates $\hat\beta$ corresponding to minimally coupled
geometry.

The solutions of Eq. (4.1) are
\begin{equation}
\Psi_{\vec b}=e^{iG_{kl}b^k\hat\beta^l}\, e^{i b_{n+1}\Phi},
\end{equation}
where $b_{n+1}=b^{n+1}$ and the quantum numbers $b^k$ ($k=1,\ldots,n+1$)
satisfy the constraint (3.42).
{}From a formal point of view it is natural to set $\hat\beta^{n+1}:=\Phi$.
Then $\Psi_{\vec b}$ with $\vec b\in \R^{n+1}$ are eigenfunctions of
the momentum operators $-i \frac{\partial}{\partial\hat\beta^k}$ with
eigenvalues $\hat p_k=b_k=G_{kl}b^l$, i.e.
\begin{equation}
-i\frac{\partial}{\partial\hat\beta^k}\Psi_{\vec b}=b_k \Psi_{\vec b},\
b_k=1,\ldots,n+1.
\end{equation}
For analogy with Sec. 3 let us investigate in more detail the degenerate
case $b^1\neq 0$, $b^2=\ldots=b^n=0$.
With
\begin{equation}
b_1=-d_1(d_1-1)b^1,\ \mbox{and}\ b_k=-d_k(d_1 b^1),\ k=2,\ldots,n
\end{equation}
we obtain
$$
b_{n+1}=b^{n+1}=\pm\left[ d_1(d_1-1) \right]^\frac{1}{2}\,b^1
	       =\mp\left[ d_1(d_1-1) \right]^{-\frac{1}{2}}\,b_1
$$
according to Eq. (3.56) from the constraint (3.42). We see
that a nonvanishing scalar field in the minimal coupling model
requires $d_1\neq 1$.

The wave function (4.2) in the degenerate case is
\begin{equation}
\Psi_{b_1,\pm}=\left[ \hat V^\frac{1}{d_1-1}\, e^{\hat\beta^1\mp\hat\Phi}
	       \right]^{i b_1},
\end{equation}
where we have defined a new minimal field coordinate
$$
\hat\Phi:=\Phi/\sqrt{d_1(d_1-1)},
$$
and
\begin{equation}
\hat V:=\prod_{k=2}^{n} \hat a_k^{\!d_k}
\end{equation}
is proportional to the volume of the internal spaces
$M_2\times\ldots\times M_n$.
For the classical solutions corresponding to the degenerate case
this quantity is constant, i.e. the internal spaces are static.

Now we show that the wave function (4.5) describes an expanding universe
if and only if $b^1>0$ (or likewise $b_1<0$).
The Hubble parameters of the spaces $M_k$, w.r.t. harmonic
time $\tau$ and for a solution (3.41) of the minimal model, are given as
\begin{equation}
\hat h_k=\frac{1}{\hat a_i}\frac{\partial \hat a_i}{\partial \tau}=b^k,\
k=1,\ldots,n.
\end{equation}
$\hat h_k>0$ corresponds to an expanding factor space $M_k$.
Since in the degenerate case only $M_1$ has nontrivial dynamics,
an expanding universe corresponds to $b^1>0$ (or likewise $b_1<0$).

The classical momenta with $\mu\equiv 1$ in the harmonic gauge are
$\hat p_k=\frac{\partial \hat L}{\partial \dot{\hat\beta}^k}=b_k$.
Thus $\hat p_1<0$ corresponds to expanding $M_1$.
Hence by Eq. (4.3) a wavefunction $\Psi_{b_1,\pm}$ describes an
expanding universe if and only if $b_1<0$ (i.e. $b^1>0$),
while the internal spaces $M_k$
with
$$
-i\frac{\partial}{\partial\hat\beta^k}\Psi_{b_1,\pm}=b_k \Psi_{b_1,\pm}
=-d_k(d_1 b^1) \Psi_{b_1,\pm}
$$
and $b^k=G^{kl}b_l=0$ for $k=2,\ldots,n$ are static.

The transformation to the conformal model has to be performed
according to Eq. (3.43) and respectively either Eq. (3.35) or Eq. (3.36),
substituting $\hat\beta^i=\beta^i+\omega(\phi)$ and $\hat\Phi(\phi)$
into Eq. (4.5), yielding
\begin{equation}
\Psi_{b_1,\pm}=\left[ e^{\frac{D-2}{d_1-1}\omega(\phi)}\,
 V^\frac{1}{d_1-1}\, e^{\beta^1\mp\hat\Phi(\phi)}
	       \right]^{i b_1},
\end{equation}
where $V:=\prod_{k=2}^{n} a_k^{\!d_k}$.

The interpretation of the wave function (4.8) in the conformal model
is complicated by the severe difficulty, that in contrast to
(4.5), which is a solution of the WdW equation (4.1), for (4.8)
it can {\em not} be expected, that it is the solution of a conformal
WdW equation.$^\bRa$
A related difficulty is that $\Psi_{b_1,\pm}$ are
only eigenfunctions of
$\frac{\partial}{\partial \hat\beta^k}=
\frac{\partial}{\partial \beta^k}
+\frac{\partial \phi}{\partial \omega}\frac{\partial}{\partial \phi}$
but not of $\frac{\partial}{\partial \beta^k}$.
As consequence the classical momenta
$p_k=\frac{\partial L}{\partial \dot\beta^k}$ cannot be obtained
as eigenvalues of a corresponding canonical operator.
In contrast to the minimal model now for all $M_k$ with $k=1,\ldots,n$
it is $p_k\not\equiv 0$, and hence also the internal factor spaces
are no longer static in the conformal model.

On classical level, the Hubble parameters of the nonsingular solution
of the conformal model w.r.t. the time parameter $\tau$ are
given as
\begin{equation}
h_k=\frac{1}{a_i}\frac{\partial a_i}{\partial \tau}=
\hat h_k+\frac{2}{D-2}\sqrt{\xi_c}b^{n+1}\tanh(\sqrt{\xi_c}b^{n+1}\tau),
\end{equation}
with $\hat h_k$ from Eq. (4.7). For expanding $M_k$ it is $h_k>0$.
In the degenerate case with $\hat h_1=b^1$ and $\hat h_k=0$
for $k=2,\ldots,n$ we yield $h_1>0$ for $\tau>\tau_0$ with $\tau_0$
given by Eq. (3.57). $M_1$ may expand for $b^1>0$ (in this case $\tau_0<0$)
as well as for $b^1<0$ (in this case $\tau_0>0$).
If $\tau>0$ the factors $M_k$ ($k=2,\ldots,n$) expand for all values of
$b_1$. This classical investigation shows that the $\Psi_{b_1,\pm}$
may describe expanding spaces $M_k$ ($k=1,\ldots,n$) for both
$b^1>0$ and $b^1<0$.

Let us perform for the degenerate case the transition
to the real Euclidean geometry, i.e. $b_1\to -i b_1$ with $\tau\to i\tau$.
Then the wave function goes from (4.5) to
\begin{equation}
\Psi_{b_1,\pm}=\left[ \hat V^\frac{1}{d_1-1}\, e^{\hat\beta^1\mp\hat\Phi}
	       \right]^{b_1}.
\end{equation}
The superposition
\begin{equation}
\Psi_{\pm}:=\sum_{b_1=0}^{\infty}\frac{(-1)^{b_1}}{b_1!}\Psi_{b_1,\pm}
\end{equation}
yields with (4.10) the wave function
\begin{equation}
\Psi_{\pm}=e^{- \hat V^\frac{1}{d_1-1}\,\hat a_1\, e^{\mp\hat\Phi} },
\end{equation}
which satisfies the quantum wormhole boundary conditions$^\HP$:
\nl
1) it is exponentially damped for large spacial geometries
(i.e. for $\hat a_1 \to \infty$ or $\hat V \to \infty$).
\nl
2) it is regular, when the spacial geometry degenerates
(i.e. when $\hat a_1 \to 0$ or $\hat V \to 0$).
\nl
Thus (4.12) may be  treated as quantum wormhole.

\np
\noindent

\section{\bf Conclusion}
\setcounter{equation}{0}
We have initially considered natural time gauges in multidimensional
universes: (i) synchronous time, (ii) conformal times of different factor
spaces, (iii) mean conformal time and (iv) harmonic time.
Transitions between them are given by special conformal coordinate
transformations. We have emphasized that conformal coordinate transformations
have to be distinguished sharply from conformal transformations of
geometrical Lagrangian models.

The conformal transformation of the minimally coupling model to the conformal
coupling
model has been performed in arbitrary dimensions $D$, with the conformal
factor and scalar field in agreement with the result of Ref. \Xa.
By Eq. (3.34) the proper generalization of the scalar field
from the conformal coupling case to that of arbitrary coupling $\xi$
has been found (Note that Eq. (5) in Ref. \Pa\ holds only for
$\xi=\xi_c=\frac{1}{6}$ with $D=4$).

We find a generalized Kasner solution for the minimally coupling model
with $M_i$ flat and zero potential, having a scale factor singularity.
Conformal transformation yields (w.r.t. harmonic time) a nonsingular
solution (3.53) for
$\phi^2<\xi^{-1}_c$ and a singular solution (3.54) for
$\phi^2>\xi^{-1}_c$.
This resolution of the scale factor singularity
of the generalized Kasner solution of the minimal model in the
corresponding conformal solution (3.53) confirms
in arbitrary dimension $D$, what has been indicated in Ref. \Ga\ for $D=4$.
At $\phi^2=\xi^{-1}_c$ there is a singularity of the conformal transformation.
The conformal equivalence between the models only holds separately in the
range $\phi^2<\xi^{-1}_c$ or $\phi^2>\xi^{-1}_c$.

In the special case of static
internal spaces in the minimal model, we find
dynamical internal spaces with a nonzero minimum  scale at $\tau=0$ for the
conformal model with external space having a minimal scale
$a_1(\tau_0)$ at (harmonic) time $\tau_0$. In the internal spaces
the conformal solution is highly asymmetric w.r.t. $\tau_0$.
Cutting the solution at $\tau_0$, the resulting pieces
allow to model the birth of universes at $\tau_0$ with different behaviour of
internal spaces in harmonic time.

The region between
$\tau_0$ and $\tau=0$ is characterized by shrinking internal spaces,
while external space expands.
However further investigations will be required to yield a more  detailed
understanding of the dynamical behaviour of internal spaces.
In Ref. \bBlZ\ first investigations for the model from Eq. (2.40)
in harmonic time gauge have shown how
the dynamics of the factor spaces $M_i$ depends critically on  the dimensions
of $M_i$.

Besides the usual quantum creation of external space $M_1$ only,
with internal spaces as spectators,
we have pointed out the possibility to create both
$M_1$ and the internal spaces by quantum tunneling from an Euclidean
region. However the initial real time is different
for the quantum creation of different factor spaces in general,
and especially in the considered model for $M_1$ and the internal
factor spaces $M_i$, $i\geq 2$. If the time delay
between creation of $M_1$ and internal spaces goes to infinity,
$\Delta\tau=\vert\tau_0\vert\to \infty$, the internal spaces remain
forever in the classically forbidden region, while external space
is given by the real Lorentzian $M_1$. Hence extra dimensions are
unobservable at any time.

Analytic continuation
of this solution to the Euclidean time region (while pertaining
real geometry) yields a purely imaginary scalar field.
This solution corresponds to an (instanton) wormhole,
where the minimal scale $a_1$ now indicates the throat of the
wormhole w.r.t. external space, and the throats of the internal spaces
are given by $\hat a_2(0), \ldots, \hat a_n(0)$.

For the minimal model we have obtained the corresponding
quantum solution from the
WdW equation.
In the degenerate case, corresponding to static internal spaces,
the solution describes a classically expanding factor $M_1$ if and only if
the classical minisuperspace momentum satisfies $\hat p_1=b_1<0$.

The wavefunction corresponding classically
to the conformal model can {\em not} be interpreted as a solution of a
conformal WdW equation.$^\bRa$ This is related to the fact that corresponding
classical momenta $p_k$ are no longer eigenvalues.
Corresponding classical internal spaces are no longer static,
while the corresponding minisuperspace momenta $p_k$ are unrestricted.
This wavefunction can describe a classically expanding $M_1$ for both
$b_1>0$ and $b_1<0$.

Performing the transition to the real Euclidean region,
a special solutions satisfying the quantum wormhole boundary
conditions$^\HP$ have been found.

\nl\nl
{\Large {\bf Acknowledgements}}
\nl\nl
This work was supported by WIP grant 016659 (U.B.),
in part by DAAD and by DFG grant 436
UKR - 17/7/93 (A. Z.) and DFG grant Bl 365/1-1 (M.R.).
A. Z. also thanks Prof. Kleinert and the Freie Universit\"at Berlin
as well as the members of the Gravitationsprojekt at
Universit\"at Potsdam for their hospitality. M. R. thanks the
Projektgruppe
Kosmologie at Universit\"at Potsdam and H.-J. Schmidt for support and
hospitality.
\newpage
\noindent
{\Large {\bf References}}
\nl\nl
$^{\aBlLP}$  U. Bleyer, D.-E. Liebscher and A. G. Polnarev,
Nuovo Cim. B {\bf 106}, 107 (1991).
\nl
$^{\bBlLP}$  U. Bleyer, D.-E. Liebscher and A. G. Polnarev,
{\em Kaluza-Klein Models},
Proc. V$^{th}$ Seminar on Quantum Gravity, Moscow 1990, World Scientific
(1991).
\nl
$^{\Iv}$  V. D. Ivashchuk Phys. Lett. A {\bf 170}, 16 (1992),
\nl
$^{\aZh}$  A. I. Zhuk, Class. Quant. Grav. {\bf 9}, 2029 (1992).
\nl
$^{\bZh}$  A. I. Zhuk, Sov J. Nucl. Phys. {\bf 55}, 149 (1992);
Phys. Rev. D {\bf 45}, 1192 (1992).
\nl
$^{\HP}$  S. W. Hawking and D. N. Page, Phys. Rev. D {\bf 42}, 2655 (1902).
\nl
$^{\Bl}$  U. Bleyer, {\em Multidimensional Cosmology}, p. 101-11 in:
{The Earth and the Universe
(A Festschrift in Honour of Hans-J\"urgen Treder)},
ed.: W. Schr\"oder, Science Ed., Bremen (1993).
\nl
$^{\IMZ}$  V. D. Ivashchuk, V. N. Melnikov, A. I. Zhuk, Nuovo Cim. B
{\bf 104}, 575 (1989),
\nl
$^{\bRa}$  M. Rainer, {\em Conformal Coupling and Invariance in Arbitrary
Dimensions}, Preprint 94/2, Mathematisches Institut,
Universit\"at Potsdam (1994).
\nl
$^{\Pa}$  D. N. Page, J. Math. Phys. {\bf 32}, 3427 (1991),
\nl
$^{\Sch}$  H.-J. Schmidt, Phys. Lett. B {\bf 214}, 519 (1988).
\nl
$^{\Gi}$  G. W. Gibbons and S. W. Hawking, Phys. Rev. D {\bf 15}, 2752 (1977).
\nl
$^{\Ma}$  K. Maeda, Phys. Rev. D {\bf 39}, 3159 (1989),
\nl
$^{\Ga}$  D. V. Gal'tsov and B. C. Xanthopoulos, J. Math. Phys. {\bf 33},
273 (1992).
\nl
$^{\Vi}$  A. Vilenkin, Phys. Rev. D {\bf 27}, 2848 (1983).
\nl
$^{\aRa}$  M. Rainer, {\em Projective Geometry for Relativistic Quantum
Physics}, Proc. 23$^{rd}$ Ann. Iranian Math. Conf. (Baktaran, 1992);
J. Math. Phys. {\bf 35}, 646 (1994).
\nl
$^{\Ze}$  H. D. Zeh, {\em The Physical Basis of the Direction of Time},
2nd. ed. Springer-Verlag (Heidelberg, 1991);
\nl
$^{\RS}$  V. A. Rubakov, M. E. Shaposhnikov, Phys. Lett. B {\bf 125},
136 (1983).
\nl
$^{\Xa}$  B. C. Xanthapoulos and Th. E. Dialynas, J. Math. Phys. {\bf 33},
1463 (1992).
\nl
$^{\bBlZ}$  U. Bleyer and A. Zhuk, {\em Multidimensional Integrable
Cosmological
Models with Dynamical and Spontaneous Compactification},
Preprint FUB-HEP/93-19, FU Berlin (1993).
\nl
\newpage

\vspace*{10truecm}
{\small Fig. 1: Quantum birth with compact Ricci flat spaces and
birth time $\tau_0\leq 0$ of external Lorentzian space $M_1$. The birth
of internal factor spaces $M_2,\ldots,M_n$ is delayed by the
interval $\Delta\tau=\vert\tau_0\vert$. For $\Delta\tau\to\infty$
the internal spaces remain for ever in the (unobservable) classically
forbidden region.}

\end{document}